\documentclass[aps, preprint]{revtex4-1}

\usepackage[utf8]{inputenc}
\usepackage{t1enc}
\usepackage{graphicx}   
\usepackage{amsmath}
\usepackage{multirow}

\begin{document}

\title{Coupled mode theory for the acoustic wave – spin wave interaction \\  
in the magphonic crystals: propagating magnetoelastic waves}
\author{Piotr Graczyk}
\email{graczyk@amu.edu.pl}
\author{Maciej Krawczyk}
\email{krawczyk@amu.edu.pl}
\affiliation{Faculty of Physics, Adam Mickiewicz University in Poznan, Umultowska 85, 61-614 Poznan, Poland}

\begin{abstract}

We have investigated co-directional and contra-directional couplings between spin wave and acoustic wave in one-dimensional periodic structure (magphonic crystal). The system consists of two ferromagnetic layers alternating in space. We have taken into consideration materials commonly used in magnonics: yttrium iron garnet, CoFeB, permalloy, and cobalt. The coupled mode theory (CMT) formalism have been successfully implemented to describe magnetoelastic interaction as a periodic perturbation in the magphonic crystal. The results of CMT calculations have been verified by more rigorous simulations by frequency-domain plane wave method and time-domain finite element method. The presented resonant coupling in the magphonic crystal is an active in-space mechanism which spatially transfers energy between propagating spin and acoustic modes, thus creating propagating magnetoelastic wave. We have shown, that CMT analysis of the magnetoelastic coupling is an useful tool to optimize and design a spin wave - acoustic wave transducer based on a magphonic crystals. The effect of spin wave damping has been included to the model to discuss the efficiency of such a device. Our model shows that it is possible to obtain forward conversion of the acoustic wave to the spin wave in case of co-directional coupling and backward conversion in case of contra-directional coupling.

\end{abstract}

\maketitle

\section{Introduction}

Coupled mode theory (CMT) is a well-known perturbation method in electromagnetism. It is used to describe coupling of modes in waveguides or Bragg reflections in periodic media \cite{Yariv2003}. In the limit of small perturbations CMT gives a simple picture of the underlying physical mechanisms, and thus it is complementary to more accurate numerical techniques like finite element method.  In particular, Solc filters, waveguide couplers and distributed feedback resonators are analytically characterized by CMT \cite{Yariv1973, Zhang2008, Sivan2016,Rashidian2003,Huang1994,Kogelnik1972}.

When two modes are resonantly coupled, the energy is exchanged between them. We are interested in wave propagation phenomena, so in this p[aper we consider the coupling-in-space mechanism only. Then, the modes may be coupled in two different ways, depending on their relative directions of the group velocities. In the case of contra-directional coupling interacting modes have opposite signs of the group velocities and the phenomenon is similar to the Bragg reflection at the frequency from the stop band where the two disperison branches of opposite slope anticrosses each other. Therefore, the mode that enters to the area where the coupling mechanism exists becomes an evanescent wave passing the energy to the other mode. The outcoming reflected mode will be almost purely the other mode, providing that the distance of interaction is long enough. On the other hand, in the co-directional coupling the modes have the same signs of the group velocities and the complete conversion of energy from one mode to another occurs periodically in space in the forward direction at the distance defined by the coupling coefficient.

In this article we present the application of CMT formalism to the interaction of spin wave and acoustic wave in the so-called magphonic crystal. Magphonic crystal is a periodic structure that is considered as a phononic and magnonic crystal simultaneously, i.e. the system that is periodic both for acoustic waves and spin waves. Multiple crossings between spin wave dispersion branches and acoustic wave dispersion branches are present as a consequence of periodicity, i.e. folding-back effect. This creates suitable conditions for co- and contra-directional couplings which does not exist in homogeneous materials.
 
The coupling mechanism between spin wave and acoustic wave is a magnetoelastic interaction (MEC), i.e. magnetostriction \cite{McKeehan1950}. The dynamic magnetization which is related to the spin wave exerts dynamic strain in the material. On the other hand, the strain associated with the acoustic wave induces dynamic effective magnetic field (inverse magnetostriction). If the strain field frequency (and its spatial distribution, i.e. wavelength) induced by the spin wave matches the frequency of magnetic effective field induced by the acoustic wave then the resonance criterion for the dynamic magnetoelastic coupling is satisfied. 

The coupled equations of motion for spin wave and acoustic wave are given in Ref. \onlinecite{Comstock1963}. It is possible to couple them in the linear regime only for acoustic waves that have a transverse component. This effect is known for an homogeneous ferromagnetic material since the fifties of XX century \cite{Kittel1958}. It is now intensively studied for the standing waves \cite{Dreher2012}, dynamic strain-mediated magnetization reversal \cite{Thevenard2016}, wave propagation \cite{Gowtham2015,Gowtham2016,Sasaki2017,Kikkawa2016,Chen2017}, as well as in the femtosecond pump-probe experiments \cite{Janusonis2016,Fahnle2017}. In Ref. \cite{Kamra2015} the transmission of acoustic energy through nonmagnetic-ferromagnetic (Pt/yttrium iron garnet) interface at the vicinity of magnetoelastic anticrossing has been investigated theoretically. However, to have complete picture of the mechanism, more rigorous dynamical description of magnetoelastic coupling in space is needed. Especially, if the goal is the effective transformation of propagating excitation from acoustic to magnetic or \textit{vice versa} then the spatial and temporal evolution of the excitations need to be investigated. 

So far, only the crossing of acoustic wave band with almost dispersionless spin wave band has been considered. This crossing acts as a band-gap for an acoustic wave but does not allow to excite propagating spin wave. Here, we show that it is possible to excite propagating magnetoelastic wave by introducing co-directional coupling at higher magnetic fields. Then, we demonstrate that by optimization of the magphonic crystal structural parameters it is possible to achieve strong co-directional coupling also for a low magnetic fields. Moreover, in the periodic system multiple contra-directional crossings are present which result in the formation of the stop bands both for spin wave and acoustic wave.

We have obtained the exact expression for coupling coefficient in the magphonic crystal for anticrossings between spin wave and acoustic wave dispersion branches from different bands. We characterized the magnitude of the coupling as a function of the mode number and filling factor for magphonic crystals composed of different materials, i.e. yttrium iron garnet (YIG), CoFeB, cobalt and permalloy, which are commonly used in magnonics. The results have been compared with the relevant dispersion relations obtained by the plane wave method (PWM). The spatial evolution of magnetoelastic modes is shown for the co-directional and contra-directional couplings. The predictions of energy transfer length from acoustic wave to spin wave obtained by CMT are verified by the time-domain simulations (finite element method). The effect of damping both in CMT and in time-domain simulations is taken into account and discussed for aforementioned phenomena. Finally, with the help of CMT and finite element simulations we optimize the magphonic crystal for bulk waves to act as a spin-acoustic wave transducer. 

\section{Coupled mode theory formalism}

\begin{figure}
\includegraphics{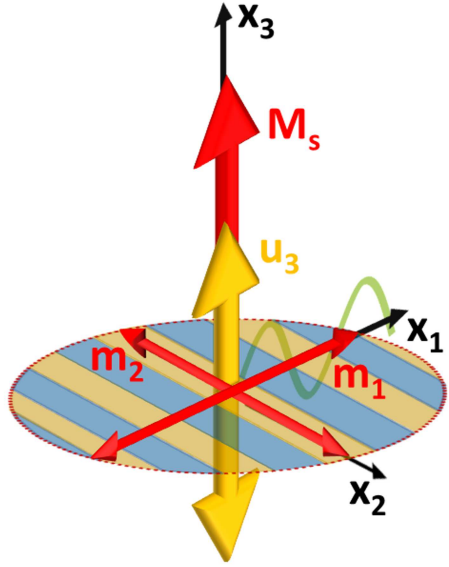}
 \caption{\label{Fig1} The geometry of considered system. The transverse acoustic wave described by displacement $u_3$ and the spin wave described by dynamic magnetization $\vec{m}=(m_1,m_2)$ propagate along $x_1$ which is the direction of periodicity of 1D magphonic structure. The external magnetic field and thus saturation magnetization $M_s$ is along $x_3$.}
\end{figure}

In the calculations the bulk spin wave with dynamic components of magnetization $m_1$ and $m_2$ (Fig. \ref{Fig1}) is coupled linearly \cite{Comstock1963} to the transverse bulk acoustic wave described by the displacement $u_3$. Both waves propagate along $x_1$ direction. The magphonic crystal consists of alternating layers of isotropic ferromagnetic materials with the periodicity along $x \equiv x_1$. 
The spin wave dynamics is described by the Landau-Lifshitz equation for magnetization components $m_1$ and $m_2$, which for one dimension and in the exchange regime takes the form:
\begin{equation}
\begin{split}
\dot{m}_1 = \omega_0 m_2 - \frac{\partial}{\partial x} \Lambda \frac{\partial m_2}{\partial x}, \\
\dot{m}_2 = - \omega_0 m_1 + \frac{\partial}{\partial x} \Lambda \frac{\partial m_1}{\partial x},
\label{eq1}
\end{split}
\end{equation}
while the acoustic wave dynamics is described by the wave equation for displacement $u\equiv u_3$:
\begin{equation}
\rho \ddot{u} = \frac{\partial}{\partial x} c \frac{\partial u}{\partial x},
\label{eq2}
\end{equation}
where $\omega_0=\gamma \mu_0 H$, $\Lambda=2A\gamma/M_s$, $\gamma=176$ GHz/T is the gyromagnetic ratio, $A$ is exchange length, $H$ is external magnetic field, $M_s$ the saturation magnetization, $\mu_0$ the magnetic susceptibility of vacuum, $\rho$ the mass density, and $c\equiv c_{44}$ is a component of the elastic tensor. The material parameters ($A$, $M_s$, $c$ and $\rho$) taken into CMT calculations for the magphonic crystal are effective parameters except magnetoelastic constant, which is periodic in space.

The solutions for homogeneous medium may be written in the form of normal modes $\widetilde{m}_{ij}\exp(k_jx-\omega t)$ and $\widetilde{u}_{j}\exp(q_jx-\omega t)$ indexed by wave vectors $k_j$, $q_j$ for a spin modes and acoustic modes respectively, at a given frequency $\omega =2\pi \nu$, $i=1,2$. The amplitudes $\widetilde{m_{ij}}$ and $\widetilde{u_j}$ are normalized here to unity. For the magphonic crystal:
\begin{equation}
\begin{split}
k_j=k_0+G_{k_j}, \\
q_j=q_0+G_{q_j},
\label{eq3}
\end{split}
\end{equation}
where $k_0$ and $q_0$ are wave vectors in the first Brillouin zone and $G_{k_j}$, $G_{q_j}$ are reciprocal lattice vectors. If $k_0=q_0$ then it is a synchronous state, otherwise it is asynchronous state.

Now we couple the equations (\ref{eq1}) and (\ref{eq2}) by magnetoelastic terms proportional to $B$:
\begin{equation}
\begin{split}
\dot{m}_1 = \omega_0 m_2 - \Lambda \frac{\partial^2 m_2}{\partial x^2}, \\
\dot{m}_2 = - \omega_0 m_1 + \Lambda \frac{\partial^2 m_1}{\partial x^2} - \gamma B \frac{\partial u}{\partial x}, \\ 
\rho \ddot{u} = c \frac{\partial^2 u}{\partial x^2}+ \frac{1}{M_s}\frac{\partial}{\partial x}B m_1.
\label{eq4}
\end{split}
\end{equation}
Then we express the solution of coupled equations as an expansion in the normal modes of the homogeneous medium:
\begin{equation}
\begin{split}
m_i=\sum_j {M_j(x)\widetilde{m}_{ij}e^{i(k_jx-\omega t)}}, \\
u=\sum_j {U_j(x)\widetilde{u}_{j}e^{i(q_jx-\omega t)}},
\label{eq5}
\end{split}
\end{equation}
where $M_j (x)$ and $U_j (x)$ are expansion coefficients which are dependent on $x$. Substituting to (\ref{eq4}), taking into account Eqs. (\ref{eq1}) and (\ref{eq2}) and neglecting second derivatives on $M_j$ and $U_j$ leads to:

\begin{widetext}

\begin{subequations}\label{eqn5}
\begin{alignat}{2}
2\Lambda \sum_j{k_j\frac{\partial M_j}{\partial x} \widetilde{m}_{1j}e^{ik_jx}}=\gamma B\sum_j{\left(q_jU_j-i\frac{\partial U_j}{\partial x} \right) \widetilde{u}_je^{iq_jx}}, \label{eqn5:a}\\
2c \sum_j{q_j\frac{\partial U_j}{\partial x} \widetilde{u}_{j}e^{iq_jx}}=\frac{1}{M_s}\sum_j{\left(iB\frac{\partial M_j}{\partial x}-\left(k_jB-i\frac{\partial B}{\partial x}\right) M_j \right) \widetilde{m}_{1j}e^{ik_jx}}. \label{eqn5:b}
\end{alignat}
\end{subequations}
Neglecting second derivatives is equivalent to the condition that $M_j(x)$ and $U_j(x)$ varies much slower in space than the wavelength of spin wave and acoustic wave (weak coupling approximation):
\begin{equation}
\begin{split}
\frac{\partial^2 U_j}{\partial x^2} \ll q \frac{\partial U_j}{\partial x}, \\
\frac{\partial^2 M_j}{\partial x^2} \ll k \frac{\partial M_j}{\partial x}.
\label{wca}
\end{split}
\end{equation}
Since the solutions of Eqs. (\ref{eq1}-\ref{eq2}) form set of orthogonal modes we can write relations:
\begin{equation}
\widetilde{u}_j \cdot \widetilde{u}_l^* = \delta_{jl},\ \widetilde{m}_{1j} \cdot \widetilde{m}_{1l}^*= \frac{1}{2}\delta_{jl}, \ \widetilde{m}_{1j}\cdot \widetilde{u}_l^* =\frac{1}{\sqrt{2}}.
\label{eq6}
\end{equation}
Using (\ref{eq6}) we multiply Eq. (\ref{eqn5:a}) by $\widetilde{m}_l^* e^{-ik_l x}$ and Eq. (\ref{eqn5:b}) by $\widetilde{u}_l^* e^{-iq_l x}$ to get:
\begin{equation}
\begin{split}
4\Lambda k_l\frac{\partial M_j}{\partial x}=\gamma B\sum_j{\left(q_jU_j-i\frac{\partial U_j}{\partial x} \right) e^{i(q_j-k_l)x}}, \\
2c q_l\frac{\partial U_j}{\partial x}=\frac{1}{M_s}\sum_j{\left(iB\frac{\partial M_j}{\partial x}-\left(k_jB-i\frac{\partial B}{\partial x}\right) M_j \right) e^{i(k_j-q_l)x}}.
\label{eq7}
\end{split}
\end{equation}
The set of differential equations (\ref{eq7}) describes coupled acoustic modes and spin modes. The strength of the coupling is expressed by magnetoelastic constant $B$ which is periodic in space in magphonic crystal. Therefore, we can expand $B$ into Fourier series with number $n$ indexing reciprocal wavenumbers $G_n$:
\begin{equation}
B=\sum_n{b_n e^{-iG_nx}}
\label{eq8}
\end{equation}
and substitute to (\ref{eq7}):
\begin{equation}
\begin{split}
4\Lambda k_l\frac{\partial M_j}{\partial x}=\gamma \sum_n{\sum_j{b_n\left(q_jU_j-i\frac{\partial U_j}{\partial x} \right) e^{i(q_j-k_l-G_n)x}}}, \\
2c q_l\frac{\partial U_j}{\partial x}=\frac{1}{M_s}\sum_n{\sum_j{b_n\left(i\frac{\partial M_j}{\partial x}-\left(k_j+G_n\right) M_j \right) e^{i(k_j-q_l+G_n)x}}}.
\label{eq9}
\end{split}
\end{equation}

The effective coupling occurs only if the phase matching criterion is satisfied. If we consider interaction between $k\equiv k_l$ and $q\equiv q_j$ modes only, then the value of $G\equiv G_n$ is fixed by the phase matching criterion to:
\begin{equation}
G=q-k \text{ and } \ n=j-l,
\label{eq10}
\end{equation}
which may be satisfied only in the synchronous state. Otherwise, in the asynchronous state, the phase difference $\Delta q=q-k-G$ appears. 

The magnetoelastic coupling is described now only by a single Fourier term $b \equiv b_n$ (Eq. \ref{eq8}) and the equations take the form (omitting subscript $j$ at the $M$ and $U$):
\begin{equation}
\begin{split}
\frac{\partial M}{\partial x}=\left(a_1U-ia_2\frac{\partial U}{\partial x} \right) e^{i\Delta qx}, \\
\frac{\partial U}{\partial x}=\left(ia_3\frac{\partial M}{\partial x}-a_4 M \right) e^{-i\Delta qx}, \\
a_1=a_2q, \ a_2=\frac{bM_s}{8Ak}, \ a_3=\frac{b}{2cM_sq}, \ a_4=a_3(k+G).
\label{eq11}
\end{split}
\end{equation}
We substitute $U\rightarrow Ue^{-iqx}$, $M\rightarrow Me^{-i(k+G)x}$ to get:
\begin{equation}
\begin{split}
\frac{\partial M}{\partial x}=\left(i(k+G)M-ia_2\frac{\partial U}{\partial x} \right), \\
\frac{\partial U}{\partial x}=\left(iqU + ia_3\frac{\partial M}{\partial x} \right).
\label{eq12}
\end{split}
\end{equation}
The general solution for a given values of $M(0)$ and $U(0)$ is:
\begin{equation}
\begin{split}
M(x)=e^{-i\overline{\beta} x} \left[\left( \cos(Dx)+i\frac{\Delta\beta}{D}\sin(Dx)\right)M(0)+\frac{\kappa_{12}}{D}\sin(Dx)U(0)\right], \\
U(x)=e^{-i\overline{\beta} x} \left[\left( \cos(Dx)-i\frac{\Delta\beta}{D}\sin(Dx)\right)U(0)-\frac{\kappa_{21}}{D}\sin(Dx)M(0)\right],
\label{eq13}
\end{split}
\end{equation}
where
\begin{equation}
\begin{split}
\overline{\beta}=\frac{G+k+q}{2}, \ \Delta \beta=\frac{G+k-q}{2}, \\
\left|\kappa_{12}\right|=\frac{bM_sq}{8Ak}, \ \left|\kappa_{21}\right|=\frac{b(k+G)}{2cqM_s}, \ \kappa^2=\kappa_{12}\kappa_{21}=\mp \frac{b}{4}\sqrt{\frac{k+G}{Ack}}, \\
D=\sqrt{\Delta\beta^2 - \kappa^2},
\label{eq14}
\end{split}
\end{equation}
where upper sign is for co-directional coupling and lower sign is for contra-directional coupling, which comes from the fact, that $\kappa_{12}$ and $\kappa_{21}$ are imaginary or real, respectively \cite{Zhang2008}. Since coupling coefficient $\kappa$ describes how fast the amplitudes $M(x)$ and $U(x)$ vary in space, the results of CMT calculations are valid under conditions (weak coupling approximation, Eq. (\ref{wca})):
\begin{equation}
\kappa \ll k \text{ and } \ \kappa \ll q.
\label{eq15}
\end{equation}
The wave vectors of coupled modes in the region close to the resonance are:
\begin{equation}
\begin{split}
\beta_1=\overline{\beta}+D, \\
\beta_2=\overline{\beta}-D.
\label{eq16}
\end{split}
\end{equation}

Dispersion relations of the coupled modes can be obtained close to the crossings by calculating new wavenumbers $\beta_1$ and $\beta_2$ from Eqs. ($\ref{eq16}$) for a given frequency $\omega$. The values of $k$ and $q$ are taken for the same frequency from the dispersion relations of uncoupled spin wave and acoustic wave, respectively.

\subsection{Co-directional coupling}
\label{sec:CoDirectionalCoupling}

The solutions of Eqs. (\ref{eq13}) for the boundary conditions $M(0)=0$ and $U(0)=U_0$, i.e. when the wave is at the starting point purely acoustic mode, are:
\begin{equation}
\begin{split}
M(x)=U_0\frac{\kappa_{12}}{D}\sin(Dx)e^{-i\overline{\beta}x}, \\
U(x)=U_0\left(\cos(Dx)-i\frac{\Delta\beta}{D}\sin(Dx)\right)e^{-i\overline{\beta}x}.
\label{eq17}
\end{split}
\end{equation}
The magnetic and acoustic modes are both sine or cosine modulated travelling waves $\exp(-i\overline{\beta}x)$.

We will consider co-directional couplings for the first bands of spin wave and acoustic wave in the homogeneous material ($j=l=n=0$) and in the magphonic crystal. Then $k=k_0$, $q=q_0$ and $G=0$. Magnetoelastic interaction is described here by the zeroth Fourier coefficient, i.e. magnetoelastic constant has a value as for an effective homogeneous medium:
\begin{equation}
b=b_0=B_1f+B_2(1-f),
\label{eq18}
\end{equation}
where $f$ is the structure filling fraction with the material 1. In the synchronous state the coupling coefficient has maximum magnitude which is:
\begin{equation}
\kappa_{\text{max}}=\frac{b}{4\sqrt{Ac}}.
\label{eq19}
\end{equation}
The energy transfer length $L_{\text{tr}}$ from acoustic mode to the spin mode is defined from zeroing cosine in Eq. (\ref{eq17}) which gives for a synchronous state:
\begin{equation}
\kappa L_{\text{tr}}=\frac{\pi}{2}.
\label{eq20}
\end{equation}
It is worth to notice that since $D$ has higher values for asynchronous state ($\Delta\beta \neq 0$ in Eq. (\ref{eq14})), the transfer of energy is more frequent in space than in synchronous state, but the total power exchanged between the modes decreases.

\subsection{Contra-directional coupling}
\label{sec:ContraDirectionalCoupling}

\begin{figure*}
\includegraphics{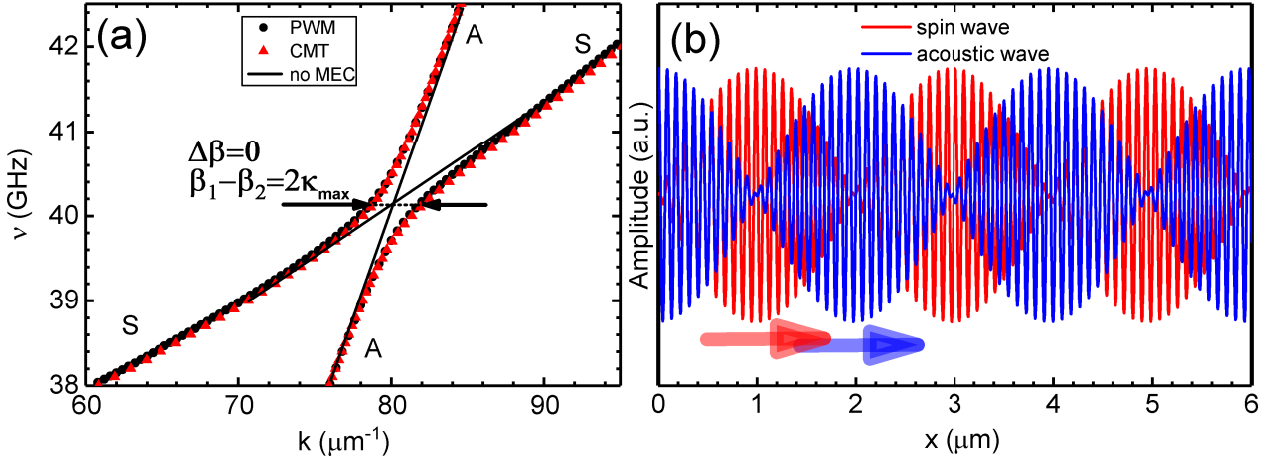}
 \caption{\label{Fig2} (a) Co-directional anticrossing of acoustic wave (A) and spin wave (S) in CoFeB calculated by CMT (red triangles) and PWM (black dots). Solid lines mark dispersion relations in the absence of magnetoelastic coupling. Arrows mark the definition of coupling coefficient $\kappa_{\text{max}}$. (b) Spatial distribution of acoustic wave displacement $u$ and spin wave dynamic magnetization $m_1$ in the co-directional coupling in CoFeB. Arrows indicate the propagation direction of waves.}
\end{figure*}

In contra-directional coupling the value of $D$ in Eq. (\ref{eq14}) becomes imaginary for the synchronous state. The solution of Eqs. (\ref{eq13}) with the substitution $T=iD$ and the boundary conditions $M(L)=0$ and $U(0)=U_0$, i.e. when the spin mode is supposed to have zero amplitude at the distance $L$ is:

\begin{equation}
\begin{split}
M(x)=U_0\frac{\kappa_{21}}{T}\frac{\cosh{TL}\sinh{ T(x-L)}-i\frac{\Delta\beta}{T}\sinh{ TL} \sinh{ T(x-L)}}{1+\frac{\kappa_{12}^2}{T^2}\sinh{TL}}e^{-i\overline{\beta}x}, \\
U(x)=U_0\frac{\cosh{Tx}-\frac{\kappa_{12}^2}{T^2}\sinh{ TL}\sinh{ T(x-L)}-i\frac{\Delta\beta}{T}\sinh{ Tx}}{1+\frac{\kappa_{12}^2}{T^2}\sinh{TL}}e^{-i\overline{\beta}x}.
\end{split}
\label{eq21}
\end{equation}
The spin and acoustic modes are both travelling waves $\exp(-i\overline{\beta}x)$ modulated by hyperbolic-sine or hyperbolic-cosine. Contra-directional coupling is possible for the periodic structure, when the folding-back effect occurs and the signs of $q$ and $k$ are opposite. Then, the coupling is described by higher Fourier coefficients of $B$ in the form:
\begin{equation}
b=b_n=\frac{B_1-B_2}{\pi n}\sin(n\pi f).
\label{eq22}
\end{equation}

\end{widetext}

\section{Results}
\label{sec:Results}

Firstly, in Sec. \ref{sec:CoDirectionalCouplingInHomogeneousMaterial} we give the description of the spin wave – acoustic wave coupling in the homogeneous ferromagnetic medium. Then, the periodicity of magnetoelastic constant in magphonic crystal that consists of alternating permalloy Ni\raisebox{-.4ex}{\scriptsize 77}Fe\raisebox{-.4ex}{\scriptsize 23}/Ni\raisebox{-.4ex}{\scriptsize 85}Fe\raisebox{-.4ex}{\scriptsize 15} (Py1/Py2) layers is introduced in Sec. \ref{sec:ContraDirectionalCouplingInPeriodicStructure} and CMT results are discussed. It is assumed that all other material parameters are the same between the layers. Next, we consider permalloy-cobalt magphonic crystal. In our model magnetoelastic constant varies periodically in space while other material parameters are taken as for effective homogeneous medium. We compare the results of the CMT with the plane wave method in frequency domain and with finite element time-domain simulations. Details of PWM calculations are given in \cite{Graczyk2017}. Finally, in Sec. \ref{sec:TimeDomainSimulationsAndTheEffectOfDamping} we discuss the effect of spin wave damping onto the multilayer optimized for acoustic wave – spin wave conversion.

\subsection{Co-directional coupling in homogeneous material}
\label{sec:CoDirectionalCouplingInHomogeneousMaterial}

 \begin{table}
\begin{tabular}{|c | c | c | c | c|} 
 \cline{2-5}
 \multicolumn{1}{c|}{} & YIG & CoFeB & Py & Co \\ [0.5ex] 
\toprule
 $M_s$ (kA/m) & 140 \cite{Stancil2009} & 1150 \cite{Conca2013} & 860 \cite{Barthelmess2004a} & 1000 \cite{Alberts1965} \\ 
 \hline
 $A$ (pJ/m) & 4 \cite{Stancil2009} & 15 \cite{Conca2013} & 13 \cite{Barthelmess2004a} & 20 \cite{Eyrich2012} \\
 \hline
 $c$ (GPa) & 76 & 70 & 50 & 80 \\
 \hline
 $\rho$ (kg/m$^3$) & 5110 & 7050 & 8720 & 8900 \\
 \hline
 $B$ (MJ/m$^3$) & 0.55 \cite{Hansen1973} & 6.5 \cite{Wang2005} & $\pm$ 0.9 \textsuperscript{\textdagger} \cite{Klokholm1981} & 10 \cite{Alberts1965} \\
\hline
$\kappa$ (1/$\mu$m) & 0.36 & 2.36 & 0.4 & 3.03 \\
\hline
$L_{\text{tr}}$ ($\mu$m) & 6.22 & 1.0 & 5.62 & 0.8 \\
\hline
$\alpha$ & 0.0003 \cite{Pirro2014} & 0.004 \cite{Conca2013} & 0.01 \cite{Rychy2016} & 0.1 \cite{Rychy2016} \\
\hline
$L_{\text{loss}}$ ($\mu$m) & 40 & 1.8 & 0.8 & 0.09 \\
 [1ex] 
 \hline
\multicolumn{5}{l}{\textsuperscript{\textdagger}\footnotesize{Minus sign for Py1, plus sign for Py2}}
\end{tabular}
\caption{Physical parameters of ferromagnetic materials and comparison of their magnetoelastic properties with the damping properties. Values of $L_{\text{loss}}$ are given for $\nu=60$ GHz.}
\end{table}

To present the co-directional mode coupling in homogeneous medium we chose CoFeB in the high magnetic field of 1 MA/m, for which the weak coupling condition (\ref{eq15}) is satisfied. Fig. \ref{Fig2}a presents dispersion relation at the vicinity of the anticrossing obtained from CMT calculations and compared with PWM method. Both results are in good agreement. Fig. \ref{Fig2}b shows the space variation of spin wave dynamic magnetization component $m_1$ and acoustic wave displacement $u$ (Eq. (\ref{eq4})) in the point of the maximum coupling (synchronous state). The acoustic wave is transformed into spin wave at a distance of about 1 $\mu$m. Then, the spin wave is transformed back to the acoustic wave. This repetitive forward transformations create propagating magnetoelastic wave.  We have obtained the same space evolution by finite element method time domain simulations (see Sec. \ref{sec:TimeDomainSimulationsAndTheEffectOfDamping}), contrary to the case of low magnetic fields in Ref. \onlinecite{Kamra2015} (and long wavelengths), where although the energy transfer occurs, the spin excitation does not propagate because of zero group velocity (flat dispersion relation) and the crossing acts effectively as a band-gap for the acoustic wave.

In Table I we correlated the parameters of four most popular materials exploited in magnonics: yttrium iron garnet (YIG), CoFeB, Py and Co. The maximum value of coupling coefficient calculated from (\ref{eq20}) is for Co and CoFeB, which give the transfer length $L_{\text{tr}}$ for that materials of about 0.8 $\mu$m and 1 $\mu$m, respectively. The transfer length is compared with the loss length $L_{\text{loss}}$, which is the distance at which the amplitude of the wave decays by the factor of $1/e$ of the initial value. In the consideration of the effect of damping we assumed, that the acoustic wave is attenuated much smaller than the spin wave and this attenuation may be neglected. It is usually correct if we compare magnetic damping coefficients with acoustic ones for shear waves in metals \cite{Mason1960,Lifshitz1999}. For bulk spin wave $L_{\text{loss}}$ is estimated from damping coefficient $\alpha$ by the formula \cite{Stancil2009}:
\begin{equation}
L_{\text{loss}}=v_g\tau=\frac{4A\gamma k}{M_s\omega \alpha}
\end{equation}
where $v_g$ is group velocity and $\tau$ is relaxation time. Thus, for the coupling-in-space mechanism, a high lifetime of a spin wave together with high group velocity is crucial. The comparison gives the conclusion that the effect of the energy transfer is completely suppressed by the spin damping in case of Co and Py. Only for CoFeB and YIG the wave has chance to transfer from spin-like to acoustic-like before being attenuated. Moreover, using high magnetic fields (in order to achieve high propagation velocities at the point of crossing as shown above for CoFeB) is somehow problematic from the point of view of applications. But the latter problem may be overcome in a periodic system.

\subsection{Contra-directional coupling in periodic structure}
\label{sec:ContraDirectionalCouplingInPeriodicStructure}

By introducing periodicity in the medium it is possible to achieve contra-directional couplings of the spin modes with the acoustic modes \cite{Graczyk2017}. We considered the propagation perpendicular to the interfaces of alternating layers (layer thickness 25 nm, $a=50$ nm, $f=0.5$) of Ni\raisebox{-.4ex}{\scriptsize 77}Fe\raisebox{-.4ex}{\scriptsize 23} (Py1) and Ni\raisebox{-.4ex}{\scriptsize 85}Fe\raisebox{-.4ex}{\scriptsize 15} (Py2) alloys. The physical parameters of Py1 are given in Table 1. We assumed that Py2 differs from Py1 only by the sign of the magnetoelastic constant \cite{Klokholm1981}. Then, the periodicity of the sample is solely due to periodicity of the magnetostriction.

\begin{figure*}
\includegraphics{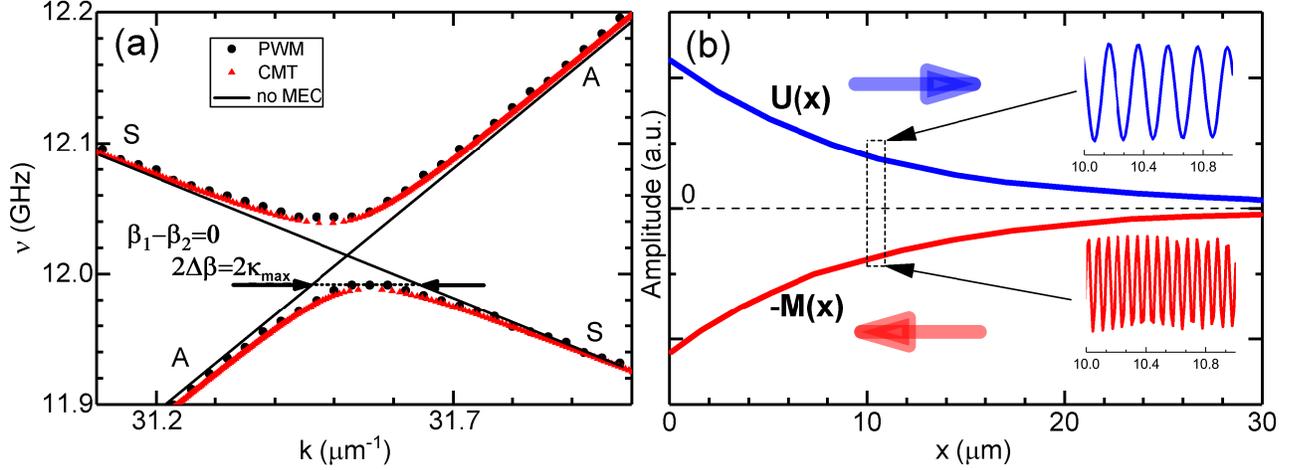}
 \caption{\label{Fig3} (a) Contra-directional C2 anticrossing of acoustic wave (A) and spin wave (S) in Py1/Py2 calculated by CMT (red triangles) and PWM (black dots). Solid lines mark dispersion relations in the absence of magnetoelastic coupling. Arrows mark the definition of coupling coefficient $\kappa_{\text{max}}$. (b) Spatial distribution of amplitudes $U(x)$ and $M(x)$ in the contra-directional coupling in Py1/Py2. Arrows indicate the propagation direction of waves. Insets show acoustic wave displacement $u$ and spin wave dynamic magnetization $m_1$ in the fragment of the structure.}
\end{figure*}

Fig. \ref{Fig3}a shows the crossing C2 (see Table II) of first acoustic mode ($j=0$) with the second spin mode ($l=-1$) of the Py1/Py2 magphonic crystal for the external magnetic field $H=100$ kA/m. Clearly, at the frequency of about 12 GHz the band gap appears. The results of CMT analytical calculations are in good agreement with PWM simulations. The amplitudes of $M(x)$ and $U(x)$ obtained from Eq. (\ref{eq21}) shows (Fig. \ref{Fig3}b) that the wave undergo Bragg reflection together with the transformation from the acoustic-like to the spin-like. The amplitude of the acoustic wave propagating into the periodic system decay exponentially, while the amplitude of the backward propagating spin wave increases. However, the distance which is needed to achieve sufficiently strong energy transfer is in the order of tens of micrometers. What is also worth to notice, is that the generated spin mode wavelength is shorter than the exciting acoustic mode, which is the result of crossing of dispersion branches of different number.

\begin{figure}
\includegraphics{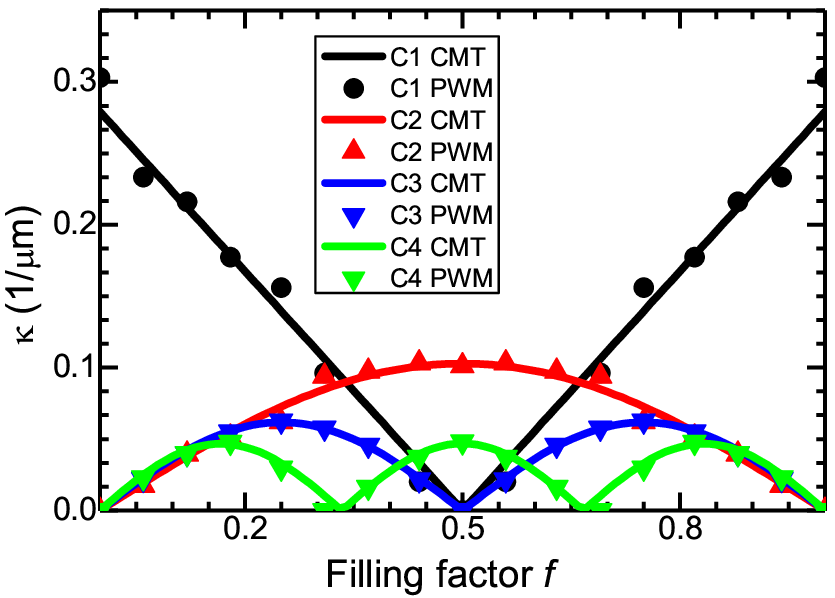}
 \caption{\label{Fig4} Dependence of coupling coefficient $\kappa$ on the filling factor $f$ in the Py1/Py2 magphonic crystal for the four different crossings. The lines are result of CMT calculations, while the points come from PWM simulations.}
\end{figure}

 \begin{table}
\begin{tabular}{|c | c | c | c | c } 
 \cline{2-4}
 \multicolumn{1}{c|}{} & $j$ & $l$ & $n=j-l$ & \multirow{5}{*}{
\includegraphics{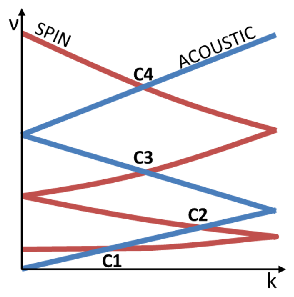}
} \\ [0.5ex] 
\cline{1-4}
 C1 & 0 & 0 & 0 & \\ 
\cline{1-4}
 C2 & 0 & -1 & 1 & \\ 
\cline{1-4}
 C3 & -1 & 1 & -2 & \\ 
\cline{1-4}
C4 & 1 & -2 & 3 & \\
 [1ex] 
\cline{1-4}
\end{tabular}
\caption{The mode numbers of acoustic wave ($j$) and spin wave ($l$) and respective number $n$ for particular crossing. }
\end{table}

The magnitude of coupling coefficient $\kappa$ has been calculated from Eq. (\ref{eq14}) in dependence on the filling factor for the four consecutive crossings labeled C1, C2, C3 and C4 (see Table 2) of the Py1/Py2 structure. The lattice constant of the structure is fixed to 50 nm and the filling factor indicates the percentage of Py2 in the system. The coupling coefficient has been determined also from the PWM dispersion relations in the way shown in Fig. \ref{Fig2}a and Fig. \ref{Fig3}a (compare with Eq. (\ref{eq16})). The results are shown in Fig. \ref{Fig4}. The coupling coefficient for the succeeding crossings obeys the relation of the succeding Fourier terms $b_n$ (Eq. (\ref{eq18}) and Eq. (\ref{eq22})) what is not surprising since $\kappa$ is proportional to the magnetoelastic constant. However, the CMT calculations fully agree with more rigorous numerical calculations, despite the fact that only the interaction of two modes are taken into account.

\begin{figure*}[t]
\includegraphics{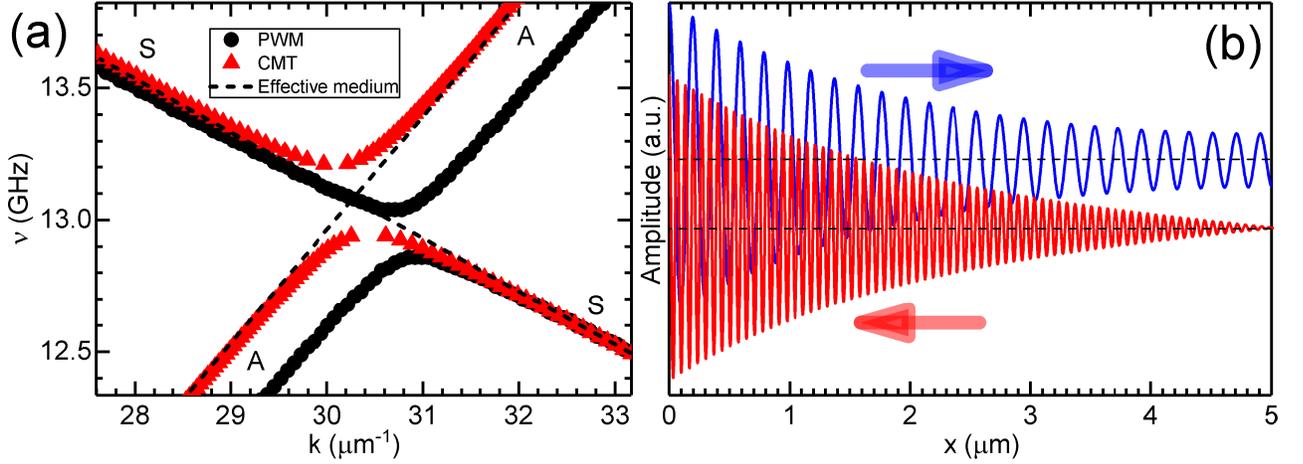}
 \caption{\label{Fig5} (a) Contra-directional C2 anticrossing of acoustic wave (A) and spin wave (S) in Py1/Co calculated by CMT (red triangles) and PWM (black dots). Dashed lines mark dispersion relations for the effective medium in the absence of magnetoelastic coupling. (b) Spatial distribution of acoustic wave displacement $u$ and spin wave dynamic magnetization $m_1$ in the contra-directional coupling in Py1/Co. Arrows indicate the propagation direction of waves.}
\end{figure*}

Next, we consider the C2 crossing of Py1/Co periodic structure of same structural parameters as previously (layer thickness 25 nm, $a=50$ nm, $f=0.5$ and magnetic field $H=100$ kA/m). The magnetoelastic coefficient vary periodically in space, while all other physical parameters ($A$, $c$, $M_s$, $\rho$) of material are calculated as for effective homogeneous medium. The comparison of C2 crossing from CMT and PWM is shown in Fig. \ref{Fig5}a. Clearly, the anticrossing calculated by CMT is shifted in wave number and frequency due to the shift of acoustic branch of effective medium with respect to the right position of acoustic branch obtained by PWM. Thus it seems that while effective medium approximation works well for the spin wave, it is not the case for acoustic waves.

In Fig. \ref{Fig5}b the Bragg reflection together with conversion from acoustic mode to spin mode and wavelength change is presented by CMT for this anticrossing. Much smaller multilayer thickness is needed for this conversion compared to previous structure (Fig. \ref{Fig3}b) as a consequence of much bigger magnetostriction in Co.

\begin{figure}
\includegraphics{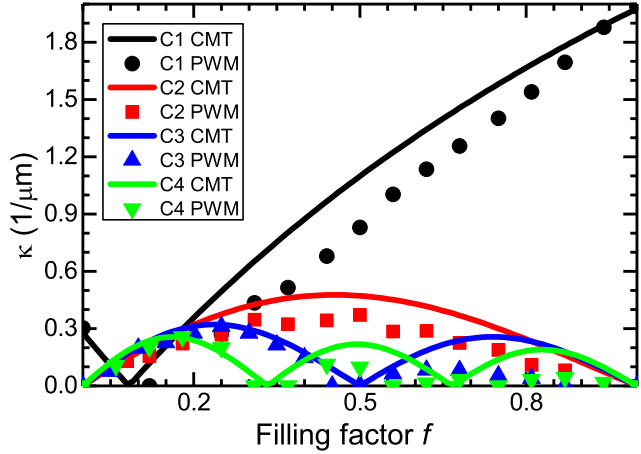}
 \caption{\label{Fig6} Dependence of coupling coefficient $\kappa$ on the filling factor $f$ in the Py1/Co magphonic crystal for the four different crossings. The lines are result of CMT calculations, while the points come from PWM simulations.}
\end{figure}

The coupling coefficient $\kappa$ for Py1/Co changes with filling factor in the similar way as for Py1/Py2 system, and is determined by $b_n$ (Fig. \ref{Fig6}). However, the maxima in C2, C3 and C4 crossings, are no more symmetrical and they have smaller high for higher filling factor (more cobalt). For example, the maximum of $\kappa$ for C2 crossing is for the $f=0.45$ what is the consequence of changes of effective material paramaters ($A$ and $c$ in Eq. [\ref{eq19}]) of the medium with the filling factor.

The comparison of CMT with PWM reveals, that although the qualitative dependence of the coupling coefficient is reproduced by CMT, it quantitatively gives overestimated values of $\kappa$, especially for high values of the filling factor. For high filling with cobalt, it seems that effective medium approximation is not accurate. However, by taking other modes into account or expanding parameters into Fourier series one can easily fit to PWM, but one also loses the most important advantage of CMT, which is its simplicity and physical transparence.

\subsection{Time-domain simulations and the effect of damping}
\label{sec:TimeDomainSimulationsAndTheEffectOfDamping}

Since we deal with coupling-in-space mechanism, it is important to consider the acoustic wave – spin wave transfer length together with the effect of damping. As mentioned in Sec. \ref{sec:CoDirectionalCouplingInHomogeneousMaterial}, the damping of spin waves in cobalt is very high and for $\alpha=0.1$ gives the loss length $L_{\text{loss}}= 90$ nm for a bulk wave with a frequency of 60 GHz. This is the distance much smaller than the transfer length (Table I), so the energy transfer due to magnetoelastic effect is completely suppressed. On the other hand, CoFeB has much lower damping together with comparable value of the magnetoelastic constant. Therefore, we solved Eqs. (\ref{eq4}) in time-domain finite-element simulations for Py1/CoFeB instead of Py1/Co multilayer structure.

\begin{figure}
\includegraphics{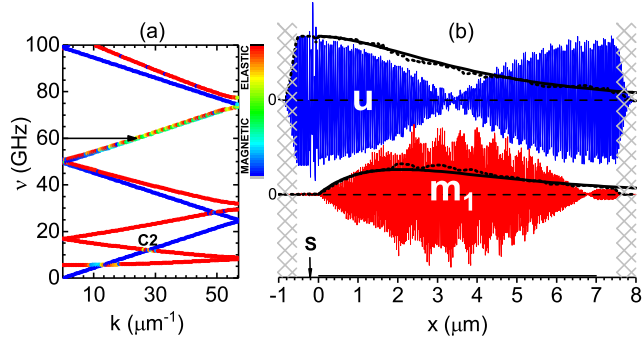}
 \caption{\label{Fig7} (a) Dispersion relation for Py1/CoFeB magphonic crystal optimized for co-directional coupling obtained by PWM simulations. (b) Spatial distribution of acoustic displacement $u$ (top) and spin wave dynamic magnetization $m_1$ (bottom) in the co-directional coupling in Py1/CoFeB. Blue and red are oscillations without damping; black dashed lines and solid lines are absolute ampitudes of the waves with damping from FEM simulations and CMT calculations, respectively.  Black horizontal line at the $x$ axis indicates the area of magphonic crystal; the acoustic wave is excited at point S; patterned areas are damping edges.}
\end{figure}

The Py1/CoFeB multilayer parameters were optimized to get broad co-directional coupling of spin and acoustic branches (Fig. \ref{Fig7}a). This was achieved for lattice constant of 55 nm and external magnetic field 160 kA/m. For that values, the third acoustic branch ($j=1$) overlaps with the fifth spin branch ($l=2$) of the dispersion relation (see Table II) and we obtain broadband magnetoelastic coupling. The details of this optimization and plane wave method calculations with magnetoelastic coupling are described in Ref. \onlinecite{Graczyk2017}. The strength of the coupling for that branches is also quite strong since the difference between spin wave number and acoustic wave number is of the one reciprocal lattice vector ($n=-1$  in Eq. (\ref{eq10})). Thus, the coupling coefficient changes with the filling factor similar as for C2 crossing (compare Table 2) in Fig. \ref{Fig4} and it gets maximum for $f=0.47$.

The transfer length for Py1/CoFeB structure is calculated by CMT from Eqs. (\ref{eq14}), (\ref{eq20}) and (\ref{eq22}) to be $L_{\text{tr}}=3.3 \mu$m. We constructed in COMSOL Multiphysics the multilayer of thickness more than twice that distance, i.e. 7 $\mu$m. The acoustic wave of 60 GHz is continuously excited at the point indicated by S (Fig. \ref{Fig7}b). Both spin and acoustic wave are damped at the edges of the simulated area to avoid reflections. The results of time-domain simulations described by Eqs. (\ref{eq4}) are shown in Fig. \ref{Fig7}b after 10 ns of excitation. Without damping, the acoustic wave is completely transformed into spin wave at the distance 3.4 $\mu$m which is almost the same as predicted by CMT calculations. The energy is transferred back to the spin wave after the distance $2L_{\text{tr}}$. It is worth to notice, that while the wavelength of acoustic wave is about 50 nm, the wavelength of the spin wave is twice smaller, about 23 nm.

The spin wave loss length is of about 1 $\mu$m in permalloy ($\alpha=0.01$) and 2 $\mu$m in CoFeB ($\alpha=0.004$) for the bulk wave at 60 GHz. Black dashed lines in Fig. \ref{Fig7}b are the absolute amplitudes of acoustic wave and spin wave with the effect of damping. From this it is visible that the acoustic wave still excite spin wave in Py1/CoFeB structure and the maximum amplitude of the spin wave is twice smaller than without damping. Then, from the distance of 3 $\mu$m the amplitude of the SW decreases but it seems that it is not transferred back into acoustic wave.

The results of the time-domain simulations are compared with the CMT calculations, where the damping effect was taken into account by introducing complex spin wave number $k=k_r+ik_i$. The value of $k_i$ is introduced to the calculations to be $k_i=1/\hat{L}_{\text{loss}}=0.8$  $\mu\text{m}^{-1}$ where $\hat{L}_{\text{loss}}$ is the averaged loss length of the Py1/CoFeB structure. Fig. \ref{Fig7} shows that the evolution of acoustic mode and spin mode amplitudes obtained by CMT are in full agreement with that of FEM simulations.

\begin{figure}
\includegraphics{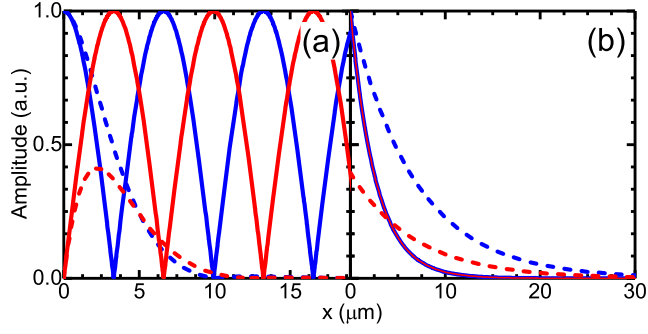}
 \caption{\label{Fig8} Amplitudes U(x) (blue line) and M(x) (red line) in Py1/CoFeB system without damping (solid lines) and with damping (dashed lines) in (a) co-directional coupling and (b) contra-directional C2 crossing.}
\end{figure}

The effect of damping onto the wave space distribution is shown again in Fig. \ref{Fig8} by comparing its amplitudes with that without damping. In case of co-directional coupling (Fig. \ref{Fig8}a) it is seen that the excited spin wave amplitude is decreased and it reaches its maximum value in a smaller distance, i.e. $x\approx 2$ $\mu$m. However, the value of $L_{\text{tr}}$ has actually increased since $U(x)$ reaches zero at $x\approx 10$ $\mu$m. Further it grows again and reaches maximum for $x\approx 12.5$ $\mu$m , while $M(x)$ reaches minimum. However, the amplitudes of both modes become neglible above $x \approx 10$ $\mu$m. 

While for a lossless system $D$, $\kappa$, $\Delta\beta$, and $\bar{\beta}$ are either real or imaginary, they become complex if the complex value of $k$ is introduced. Then, neither $U(x)$ nor $M(x)$ may be considered as a functions described by purely trigonometric or purely hyperbolic functions, as in case of co-directional and contra-directional couplings described in Secs. \ref{sec:CoDirectionalCoupling} and \ref{sec:ContraDirectionalCoupling}. They are complex superposition of trigonometric functions describing mode energy transfer and exponential decay due to a damping. Therefore, the maximum value of $M(x)$ is shifted to a smaller distance because of the exponential damping of the wave, despite the fact of higher value of transfer length. It is worth to underline, that while in the case of asynchronous state the $L_{\text{tr}}$ becomes smaller but the energy exchange is only partial, damping causes the increase of $L_{\text{tr}}$ but modes exchange all amount of the energy which is not yet lost.

For completeness, in Fig. \ref{Fig8}b we present $M(x)$ and $U(x)$ in the lossless and damped Py1/CoFeB system for the contra-directional crossing C2 (Fig. \ref{Fig7}a). It is seen, that the distance required for almost full power exchange between modes is larger for the damped system and the outgoing spin wave has more than twice reduced amplitude.

\section{Summary}

The co-directional and contra-directional couplings between spin waves and acoustic waves in the magphonic crystal have been described. It is an active in-space mechanism which transfers energy between magnetic and mechanic degrees of freedom. Coupled mode theory formalism allowed for quantitative description of the MEC strength. The structure has been optimized for an efficient and broadband co-directional coupling. In this case magnetoelastic wave propagates through the magphonic crystal. The phenomenon may be utilized for a conversion of acoustic wave to the spin wave or \textit{vice-versa}. For example, the Py1/CoFeB multilayer considered above should be of about 2 $\mu$m thick to obtain maximum forward energy transfer to the spin wave at the output (Fig. \ref{Fig8}a). On the other hand, in contra-directional coupling the magphonic crystal thickness should be as thick as possible to obtain backward spin wave at the output (in case of acoustic wave at the input).

We have shown that it is possible to resonantly excite spin wave by acoustic wave in the case of bulk waves, where the damping is very strong. From Table I it is evident that YIG is the most promising candidate in the context of magnetoelastic coupling. The value of the transfer length is much less than the loss length. Furthermore, it should be possible to achieve this effect in thin films, since the loss lengths are much larger for magnetostatic surface and volume spin waves than for bulk exchange waves. Further engineering of the band structures in magphonic crystals together with the development of CMT analysis for surface waves are required to optimize this effect.

\begin{acknowledgments}
We would like to thank Jarosław Kłos for valuable remarks during preparation of the manuscript.
The study has received financial support from the National Science Centre of Poland under grants UMO-2012/07/E/ST3/00538, UMO-2016/21/B/ST3/00452 and the EU’s Horizon 2020 research and innovation programme under Marie Sklodowska-Curie GA No.~644348 (MagIC).
\end{acknowledgments}
%

\begin{thebibliography}{10}

\bibitem{Yariv2003}
A.~Yariv and P.~Yeh, {\em {Optical Waves in Crystals}}.
\newblock John Wiley \& Sons, 2003.

\bibitem{Yariv1973}
A.~Yariv, ``{Coupled-Mode theory for guided-wave optics},'' {\em IEEE Journal
  of Quantum Electronics}, vol.~9, no.~9, pp.~919--933, 1973.

\bibitem{Zhang2008}
K.~Zhang and D.~Li, {\em {Electromagnetic Theory for Microwaves and
  Optoelectronics}}.
\newblock Berlin: Springer-Verlag, 2008.

\bibitem{Sivan2016}
Y.~Sivan, S.~Rozenberg, and A.~Halstuch, ``{Coupled-mode theory for
  electromagnetic pulse propagation in dispersive media undergoing a
  spatiotemporal perturbation: Exact derivation, numerical validation, and
  peculiar wave mixing},'' {\em Physical Review B - Condensed Matter and
  Materials Physics}, vol.~93, no.~14, p.~144303, 2016.

\bibitem{Rashidian2003}
B.~Rashidian and S.~Khorasani, ``{Coupled Mode Theory of Waveguides with
  Conducting Interfaces},'' 2003.

\bibitem{Huang1994}
W.-P. Huang, ``{Coupled-mode theory for optical waveguides: an overview},''
  {\em Journal of the Optical Society of America}, vol.~11, no.~3,
  pp.~963--983, 1994.

\bibitem{Kogelnik1972}
H.~Kogelnik and C.~V. Shank, ``{Coupled-wave theory of distributed feedback
  lasers},'' {\em Journal of Applied Physics}, vol.~43, no.~5, pp.~2327--2335,
  1972.

\bibitem{McKeehan1950}
C.~Kittel, ``{Physical theory of ferromagnetic domains},'' {\em Reviews of
  Modern Physics}, vol.~21, no.~4, pp.~555--558, 1949.

\bibitem{Comstock1963}
R.~L. Comstock and B.~A. Auld, ``{Parametric Coupling of the Magnetization and
  Strain in a Ferrimagnet. I. Parametric Excitation of Magnetostatic and
  Elastic Modes},'' {\em Journal of Applied Physics}, vol.~34, no.~5,
  pp.~1461--1464, 1963.

\bibitem{Kittel1958}
C.~Kittel, ``{Interaction of Spin Waves and Ultrasonic Waves in Ferromagnetic
  Crystals},'' {\em Physical Review}, vol.~110, no.~4, pp.~836--841, 1958.

\bibitem{Dreher2012}
L.~Dreher, M.~Weiler, M.~Pernpeintner, H.~Huebl, R.~Gross, M.~Brandt, and
  S.~Goennenwein, ``{Surface acoustic wave driven ferromagnetic resonance in
  nickel thin films: Theory and experiment},'' {\em Physical Review B},
  vol.~86, no.~13, p.~134415, 2012.

\bibitem{Thevenard2016}
L.~Thevenard, I.~S. Camara, S.~Majrab, M.~Bernard, P.~Rovillain,
  A.~Lema{\^{i}}tre, C.~Gourdon, and J.-Y. Duquesne, ``{Precessional
  magnetization switching by a surface acoustic wave},'' {\em Physical Review
  B}, vol.~93, no.~13, p.~134430, 2016.

\bibitem{Gowtham2015}
P.~G. Gowtham, T.~Moriyama, D.~C. Ralph, and R.~A. Buhrman, ``{Traveling
  surface spin-wave resonance spectroscopy using surface acoustic waves},''
  {\em Journal of Applied Physics}, vol.~118, no.~23, 2015.

\bibitem{Gowtham2016}
P.~G. Gowtham, D.~Labanowski, and S.~Salahuddin, ``{The mechanical back-action
  of a spin-wave resonance in a magnetoelastic thin film on a surface acoustic
  wave},'' {\em Physical Review B}, vol.~94, p.~014436, 2016.

\bibitem{Sasaki2017}
R.~Sasaki, Y.~Nii, Y.~Iguchi, and Y.~Onose, ``{Nonreciprocal propagation of
  surface acoustic wave in Ni / LiNbO 3},'' {\em Physical Review B}, vol.~95,
  p.~020407(R), 2017.

\bibitem{Kikkawa2016}
T.~Kikkawa, K.~Shen, B.~Flebus, R.~A. Duine, K.-I. Uchida, Z.~Qiu, G.~E.~W.
  Bauer, and E.~Saitoh, ``{Magnon Polarons in the Spin Seebeck Effect},'' {\em
  Physical Review Letters}, vol.~117, p.~207203, 2016.

\bibitem{Chen2017}
C.~Chen, A.~Barra, A.~Mal, G.~Carman, and A.~Sepulveda, ``{Voltage induced
  mechanical / spin wave propagation over long distances},'' {\em Applied
  Physics Letters}, vol.~110, p.~072401, 2017.

\bibitem{Janusonis2016}
J.~Janu{\v{s}}onis, T.~Jansma, C.~L. Chang, Q.~Liu, A.~Gatilova, and A.~M.
  Lomonosov, ``{Transient Grating Spectroscopy in Magnetic Thin Films :
  Simultaneous Detection of Elastic and Magnetic Dynamics},'' {\em Scientific
  Reports}, vol.~6, p.~29143, 2016.

\bibitem{Fahnle2017}
M.~F{\"{a}}hnle, T.~Tsatsoulis, C.~Illg, M.~Haag, B.~Y. M{\"{u}}ller, and
  L.~Zhang, ``{Ultrafast Demagnetization After Femtosecond Laser Pulses:
  Transfer of Angular Momentum from the Electronic System to Magnetoelastic
  Spin-Phonon Modes},'' {\em Journal of Superconductivity and Novel Magnetism},
  pp.~1--7, 2017.

\bibitem{Kamra2015}
A.~Kamra, H.~Keshtgar, P.~Yan, and G.~E.~W. Bauer, ``{Coherent elastic
  excitation of spin waves},'' {\em Physical Review B}, vol.~91, p.~104409,
  2015.

\bibitem{Graczyk2017}
P.~Graczyk, J.~K{\l}os, and M.~Krawczyk, ``{Broadband magnetoelastic coupling
  in magnonic-phononic crystals for high-frequency nanoscale spin-wave
  generation},'' {\em Physical Review B}, vol.~95, no.~10, p.~104425, 2017.

\bibitem{Stancil2009}
D.~Stancil and A.~Prabhakar, {\em {Spin Waves}}.
\newblock Springer, 2009.

\bibitem{Conca2013}
A.~Conca, J.~Greser, T.~Sebastian, S.~Klingler, B.~Obry, B.~Leven, and
  B.~Hillebrands, ``{Low spin-wave damping in amorphous Co40Fe40B 20 thin
  films},'' {\em Journal of Applied Physics}, vol.~113, no.~21, 2013.

\bibitem{Barthelmess2004a}
M.~Barthelmess, C.~Pels, a.~Thieme, and G.~Meier, ``{Stray fields of domains in
  permalloy microstructures - Measurements and simulations},'' {\em Journal of
  Applied Physics}, vol.~95, no.~2004, pp.~5641--5645, 2004.

\bibitem{Alberts1965}
H.~Alberts and L.~Alberts, ``{On the magnetization and magnetostriction of
  polycrystalline cobalt},'' {\em Physica}, vol.~31, no.~7, pp.~1063--1068,
  1965.

\bibitem{Eyrich2012}
C.~Eyrich, W.~Huttema, M.~Arora, E.~Montoya, F.~Rashidi, C.~Burrowes,
  B.~Kardasz, E.~Girt, B.~Heinrich, O.~N. Mryasov, M.~From, and O.~Karis,
  ``{Exchange stiffness in thin film Co alloys},'' {\em Journal of Applied
  Physics}, vol.~111, no.~7, pp.~2012--2015, 2012.

\bibitem{Hansen1973}
P.~Hansen, ``{Magnetostriction of ruthenium-substituted yttrium iron garnet},''
  {\em Phys. Rev. B}, vol.~8, no.~1, p.~246, 1973.

\bibitem{Wang2005}
D.~Wang, C.~Nordman, Z.~Qian, J.~M. Daughton, and J.~Myers, ``{Magnetostriction
  effect of amorphous CoFeB thin films and application in spin-dependent tunnel
  junctions},'' {\em Journal of Applied Physics}, vol.~97, no.~10, p.~10C906,
  2005.

\bibitem{Klokholm1981}
E.~Klokholm and J.~A. Aboaf, ``{The saturation magnetostriction of permalloy
  films},'' {\em Journal of Applied Physics}, vol.~52, pp.~2474--2476, 1981.

\bibitem{Pirro2014}
P.~Pirro, T.~Br{\"{a}}cher, A.~V. Chumak, B.~L{\"{a}}gel, C.~Dubs,
  O.~Surzhenko, P.~G{\"{o}}rnert, B.~Leven, and B.~Hillebrands, ``{Spin-wave
  excitation and propagation in microstructured waveguides of yttrium iron
  garnet/Pt bilayers},'' {\em Applied Physics Letters}, vol.~104, no.~1,
  pp.~10--14, 2014.

\bibitem{Rychy2016}
J.~Rych{\l}y, J.~W. K{\l}os, and M.~Krawczyk, ``{Spin wave damping in periodic
  and quasiperiodic magnonic structures},'' {\em Journal of Physics D: Applied
  Physics}, vol.~49, no.~17, p.~175001, 2016.

\bibitem{Mason1960}
W.~P. Mason, ``{Phonon Viscosity and its effect on acoustic wave attenuation
  and dislocation motion},'' {\em The Journal of the Acoustical Society of
  America}, vol.~32, no.~4, pp.~458--472, 1960.

\bibitem{Lifshitz1999}
R.~Lifshitz and M.~L. Roukes, ``{Thermoelastic Damping in Micro- and
  Nano-Mechanical Systems},'' {\em Phys. Rev. B}, vol.~61, no.~8, p.~5600,
  1999.

\end{thebibliography}
%

\end{document}